\documentclass[aps,prl,secnumarabic,twocolumn,
groupedaddress,showpacs,amsmath,amssymb]{revtex4}
\usepackage[cp1251]{inputenc}

\usepackage{bm}
\usepackage{graphicx}

\begin{document}

\title{Condensation of phonons in an ultracold Bose gas}
\author{Yu. Kagan and L.A. Manakova}
\email[]{manakova@kurm.polyn.kiae.su} \affiliation{ RRC "Kurchatov Institute", 123182
Kurchatov Sq.,1, Moscow, Russia}

\begin{abstract}
We consider the generation of longitudinal phonons in an elongated
Bose-condensed gas at zero temperature due to parametric resonance
as a result of the modulation of the transverse trap frequency. The
nonlinear temporal evolution with account of the phonon-phonon
interaction leads self-consistently to the formation of the
stationary state with the macroscopic occupation of a single phonon
quantum state.
\end{abstract}

\maketitle

The problem of the Bose-Einstein condensation of excitations is one
of most interesting aspects of the condensed matter physics. It is
well known that under condition of thermal equilibrium  the chemical
potential of excitations vanishes and, as a result, their condensate
does not form. The only way to overcome this obstacle is a creation
of the stationary state with the conserved number of excitations in
a non-equilibrium system. In the present paper we describe a
mechanism leading to the formation of such state for phonon
excitations in the Bose-condensed atomic gases. The mechanism is
based on generating  phonons with the help of parametric resonance.
In the Bose-condensed gases the parametric resonance (PR) is
responsible for the effective energy transfer between different
branches of excitations and temporal evolution of the system. In
particular, the damping of transverse oscillations in a elongated
cylindrical trap at zero temperature is due to PR with the
production of longitudinal phonons \cite{K1}. The nontrivial picture
of the damping in these conditions is revealed in the remarkable
work of the Paris group \cite{Ch}. The specific features of this
picture is a result of the nonlinear evolution accompanying PR in
such system \cite{K2}.

We analyze the temporal evolution of interacting longitudinal
(sound) phonons in an elongated cylindrical parabolic trap subjected
to permanent modulation of the transverse trap frequency
$\omega_{\bot}$. This modulation brings about the oscillations of
the condensate density and, accordingly, sound velocity. Due to PR
this results in the generation of pairs of sound phonons with the
opposite momenta and energies close to a half of the modulation
frequency $\omega_0/2$. In fact, phonon pairs are produced within
the finite energy interval $\omega_0/2\pm E_0$. The parameter $E_0$
is connected with the modulation amplitude
$|\delta\omega_{\bot}/\omega_{\bot}|\ll 1$ by the relation $E_0\sim
\omega_0\cdot|\delta\omega_{\bot}/\omega_{\bot}|$. If $E_0$ exceeds
the damping factor $\gamma$ for longitudinal phonons, their
occupation numbers begin to increase exponentially in time. In this
case the phonon-phonon interactions become significant. As a result,
the evolution scenario changes drastically. Let us assume that, due
to finite value of longitudinal size L, a single phonon level alone
lies within the narrow energy interval $2E_0$. In this case the main
effect of the phonon-phonon interactions is an effective
renormalization of the level. As is shown below, in the course of
the self-consistent nonlinear evolution the renormalized level
reaches the left or right edge of the parametric resonance interval
in which the growth of the phonon occupation numbers ceases. As a
result, the stationary state of phonons with the macroscopic
occupation numbers in a single quantum state takes place. In the new
state the steady periodic space modulation of the longitudinal
density of a Bose gas occurs  with the amplitude proportional to the
number of condensed phonons.

Let us consider the Bose-condensed gas at $T=0$ in a trap of the
cylindrical symmetry with $L\gg R$, where $R$ is the radius of the
condensate. Neglecting the edge effects, we can write the general
equation for the field operator $\hat{\Psi}({\bf r},z,t)$ of atoms
in the form
\begin{equation} \label{eq:1}
i\hbar\frac{\partial \hat{\Psi}}{\partial
t}\!\!=\!\!\Big[\!\!-\frac{\hbar^2}{2m}\nabla^2_{{\bf
r}}-\frac{\hbar^2}{2m}\frac{\partial^2}{\partial z^2}+\frac{m\omega^2(t){\bf
r}^2}{2}\Big]\hat{\Psi}+U_0 \hat{\Psi}^+\hat{\Psi}\hat{\Psi},
\end{equation}
where $U_0=4\pi\hbar^2 a/m$, $a$ being the s-scattering length. Note, considering a
dilute and cold Bose gas with dominant binary collisions we imply as usually that the
radius of interaction region $r_0\ll a$, the gas parameter $na^3\ll 1$, and the
correlation length $\xi> n^{-1/3}$, $n$ is the atomic density. In this case, the
replacement of two-body interaction with the effective contact potential is relevant.
The trap transverse frequency depends on time as
\begin{equation} \label{eq:2}
\omega(t)=\omega_{\bot}(1+\eta\sin\omega_0 t),\;\;\;\eta\ll 1.
\end{equation}
According to the results \cite{K3}, we introduce the spatial
scaling parameter $b(t)$ and new variables
$\overrightarrow{\rho}={\bf r}/b(t)$, $\tau(t)$. The field
operator in terms of new variables can be written as
\begin{equation} \label{eq:3}
\hat{\Psi}=\frac{\hat{\chi}(\overrightarrow{\rho},z,\tau)}{b(t)}\cdot
e^{i\Phi};\;\;\;\;\Phi=\frac{mr^2}{2\hbar}\cdot\frac{\dot{b}}{b}.
\end{equation}
Substituting these expressions into Eq.(\ref{eq:1}), we obtain
straightforwardly
\begin{equation} \label{eq:4}
i\hbar\frac{\partial \hat{\chi}}{\partial
\tau}\!\!=\!\!\Big[\!\!-\frac{\hbar^2}{2m}\nabla^2_{{\bf
\rho}}\!+\!\frac{m\omega^2_{\bot}{\bf
\rho}^2}{2}\Big]\hat{\chi}\!+\!U_0
\hat{\chi}^+\hat{\chi}\hat{\chi}\!-\!b^2(t)\frac{\hbar^2
}{2m}\frac{\partial^2\hat{\chi}}{\partial z^2},
\end{equation}
provided that functions $b(t)$ and $\tau(t)$ satisfy the equations
\begin{equation} \label{eq:5}
\frac{d^2 b}{dt^2}+\omega^2(t)b=\frac{\omega^2_{\bot}}{b^3};\;\;\;\;
b^2\frac{d\tau}{dt}=1.
\end{equation}
Treating the condensate wave function as independent of $z$, we find
that Eq.(\ref{eq:4}) expressed in variables  ${\bf \rho}, \tau$
describes the evolution at the time-independent frequency
$\omega_{\bot}$. Combining expression (\ref{eq:2}) with the
condition $\omega_0\ll\omega_{\bot}$, we find the solution of Eq.
(\ref{eq:5}) in the form
\begin{equation} \label{eq:6}
b(t)=1+b_1(t),\;\;\;\;\; b_1(t)\approx \frac{-\eta}{2}\cdot
\sin\omega_0 t.
\end{equation}
The field operator $\hat{\chi}$ can be represented in the usual
form $\hat{\chi}=(\chi_0+\hat{\chi}')\exp(-i\mu\tau)$, $\chi_0$
being the condensate wave function and $\mu$ being the initial
chemical potential. In the absence of excitations at $T=0$ we can
omit the last term in Eq. (\ref{eq:4}) and  determine
$\chi_0(\rho)$ in the universal form. Considering excited states,
we first linearize Eq. (\ref{eq:4}) in $\hat{\chi}'$. Under
condition $\omega_0\ll\omega_{\bot}$ only long-wave longitudinal
phonons prove to be involved into the evolution of the system. For
these phonons, the transverse distribution is close to $\chi_0$
(see, e.g., \cite{Z}). This allows us to simplify the equation
obtained. Using the second relation in (\ref{eq:5}) and going over
to variable $t$, we arrive at the following equation
\begin{equation} \label{eq:10a}
i\hbar\frac{\partial \hat{\chi}'}{\partial
t}=\frac{G}{b^2(t)}(\hat{\chi}'+\hat{\chi}^{'+})-\frac{\hbar^2
}{2m}\frac{\partial^2\hat{\chi}'}{\partial z^2};\;\;G=U_0 \chi_0^2,
\end{equation}
Involving the symmetry of the problem,  we have for the operator
$\hat{\chi}'$ in second quantization
\begin{equation} \label{eq:11}
\hat{\chi}'=\sum\limits_k \chi_k
\hat{a}_k;\;\;\;\;\chi_k=\frac{e^{ikz}}{\sqrt{L}}\cdot\phi_0(\overrightarrow{
\rho}),
\end{equation}
where $\hat{a}_k$ being the annihilation operator for atoms. At
$\omega_0\ll\omega_{\bot}$ the function $\phi_0$ is actully close to $\chi_0$ for both
the quasi-1D and Thomas-Fermi cases. In the expression (\ref{eq:11}) we keep the
ground state alone for the transverse motion assuming that higher states are
insignificant in the course of evolution. Substitution of the expression (\ref{eq:11})
into Eq.(\ref{eq:10a}) gives after the standard averaging over the radial variables
\begin{equation} \label{eq:12}
i\hbar\frac{d\hat{a}_k}{d t}=\Big(\bar{G}+\frac{\hbar^2
k^2}{2m}\Big)\hat{a}_k+\bar{G}\hat{a}_{-k}^+-\eta \bar{G}\sin\omega_0
t(\hat{a}_k+\hat{a}_{-k}^+)
\end{equation}
$\bar{G}=\int d^2
\rho\phi_0^2(\overrightarrow{\rho})G(\overrightarrow{\rho})$. Let
us rewrite this equation in terms of phonon operators using the
usual transformation $\hat{a}_k=u_k \hat{b}_k+v_k \hat{b}_{-k}^+$.
When the last term in Eq. (\ref{eq:12}) is omitted, a set of
equations for $\hat{b}_k,\hat{b}_{-k}^+$ determines the well-known
Bogoliubov spectrum $\omega_k$ (with the replacement $
nU_0\rightarrow \bar{G}$). Considering the general nonstationary
case, we introduce the substitution $\hat{b}_k=\hat{\tilde{b}}_k
\exp(-i\omega_0 t/2)$ and take into account that long times
$\omega_0 t\gg 1$ are most interesting for the analysis. Then,
within the resonance approximation the equations for
$\hat{b}_k,\hat{b}_{-k}^+$ are reduced to the form
\begin{equation} \label{eq:14}
i\frac{d\hat{\tilde{b}}_k}{dt}=\xi_k\hat{\tilde{b}}+iE_{0k}
\hat{\tilde{b}}^+_{-k};\;\;\;-i\frac{d\hat{\tilde{b}}^+_{-k}}{dt}=\xi_k\hat{\tilde{b}}^+_{-k}-iE_{0k}
\hat{\tilde{b}},
\end{equation}
$E_{0k}=(\eta\bar{G}/2\hbar)[(u_k+v_k)/u_k]$,
$\xi_k=\omega_k-\omega_0/2$. In fact, we consider the evolution of
excitations at the background of the coherently oscillating
condensate. These excitations can be found as oscillations of
classical Bose-field of the condensate (see, e.g., \cite{P}). With
this advantage the operators in Eqs. (\ref{eq:14}) can be replaced
with the classical functions. If the solution of Eqs. (\ref{eq:14})
is represented in the form $\tilde{b}_k,\tilde{b}_{-k}^+\sim
\exp(\alpha_{0k} t)$, one finds
$\alpha_{0k}=[|E_{0k}|^2-\xi_k^2]^{1/2}$. This result demonstrates
the appearance of PR with an exponential growth of the phonon
occupation numbers at $|E_0|>|\xi_k|$, which is induced by the
modulation of the transverse trap frequency. Thus the parametric
resonance occurs within the narrow range near $\omega_0$ with the
width $2E_0$. For sound phonons
$|(u_k+v_k)/u_k|\approx\hbar\omega_k/\mu$ and for the parameter
$E_0$, we find
\begin{equation} \label{eq:15}
E_{0k}\approx \eta\omega_0=\omega_0\frac{|\delta \omega_{\bot}|}{\omega_{\bot}}\equiv
E_{0}.
\end{equation}

So far we have disregarded the phonon-phonon interaction. With an
exponential growth of the phonon number the interaction begins to
play an essential role. The weakness of the phonon-phonon
interaction as itself (see below) allows us to take only three- and
four-phonon processes into account. Assuming generation of sound
phonons alone  in the parametric interval $2E_0$, we can use the
expressions for $H^{(3)}$ and $H^{(4)}$ obtained within the
hydrodynamic approximation, see \cite{L}. The direct analysis shows
that the contribution of $H^{(4)}$ is small in comparison with the
term calculated in second order of the perturbation theory with
respect to $H^{(3)}$ owing to smallness of the gas parameter. The
dominant term in the Hamiltonian of three-phonon interactions has
the form
\begin{equation} \label{eq:16}
H^{(3)}= \frac{m}{2}\int d^3x \hat{{\bf v}}\delta\hat{n}\hat{{\bf
v}}.
\end{equation}
Here $\hat{{\bf v}}$, $\delta\hat{n}$ are the operators of
velocity and alternating part of the density, respectively. In
fact, we consider the interaction of sound phonons that reduces
$H^{(3)}$ to the one-dimensional problem. Since each three-phonon
vertex implies the momentum conservation law, at least one of
three phonons lies beyond the parametric interval and, therefore,
has zero occupation number at $T=0$. This makes possible to reduce
the expression obtained in second order in $H^{(3)}$ to the
effective Hamiltonian for the four-phonon interaction ($\Delta(k)$
is the Kroneker symbol)
\begin{equation} \label{eq:17}
H_{eff}=\frac{\hbar A}{2}\sum\limits_{k_1,...,k_4}
\hat{b}^+_{k_1}\hat{b}^+_{k_2}\hat{b}_{k_3}\hat{b}_{k_4}\Delta(k_1+k_2-k_3-k_4).
\end{equation}
Here all states lie within the interval of $2E_0$. Using the known
expressions for operators $\hat{{\bf v}}$, $\delta \hat{n}$ \cite{L}
and condition $E_0\ll \omega_0$, we obtain $A=\hbar \omega_0^2/\mu
N$ where $N$ is the total number of particles. In addition, second
order in $H^{(3)}$ contains the imaginary part related to real decay
processes of phonons, which determine the phonon damping. Later, we
take into account phenomenologically introducing a decrement
$\gamma_k$. After  work \cite{Ch} one can conclude that the
parameter $\gamma_k$ is small at $T=0$ for the geometry under
consideration. Therefore, it is rather easy to satisfy the
conditions when $\alpha_{0k}>\gamma_k$ and the parametric growth of
phonon number remains. Now let us write the equation for $\hat{b}_k$
taking $H_{eff}$ into account. Using substitution
$\hat{b}_k=\hat{\tilde{b_k}} \exp(-i\omega_0 t/2)$ and going over to
the classical Bose field for phonons, we have
\begin{equation} \label{eq:19}
\begin{split}
i\frac{d\tilde{b}_k}{dt}=&(-i\gamma_k+\xi_k)\tilde{b}_k+iE_0
\tilde{b}^*_{-k}+\\&+A\sum\limits_{k_2k_3k_4}
\tilde{b}^*_{k_2}\tilde{b}_{k_3}\tilde{b}_{k_4}\Delta(k+k_2-k_3-k_4).
\end{split}
\end{equation}
By means of the equations for $\tilde{b}_{k}$ and $\tilde{b}^*_{-k}$
one can directly obtain the equations for the correlators
$N_k=<\tilde{b}^+_k \tilde{b}_k>$ and $f_k=<\tilde{b}_k
\tilde{b}_{-k}>$. Using conditions $A\ll E_0\ll\omega_0$, we
decouple the  arising four-phonon terms within the mean-field
approximation. As a result, we arrive at a set of nonlinear
equations that describes  the self-consistent evolution of
interacting phonons in the vicinity of PR
\begin{equation} \label{eq:20}
\begin{split}
&\;\;\;\frac{dN_k}{dt}\!\!=\!\!\!-2\gamma N_k\!+\!E_{0}
(f_k\!+\!f^*_k)\!+\!iA(\mathcal{P}^{*}\!f_k\!-\!\! \mathcal{P}
f^*_k);\\&i\frac{df_k}{dt}\!\!=\!\!2(\!\!-i\gamma\!+\!\xi_k)f_k\!+\!2iE_0
N_k\!+\!2A\mathcal{P} N_k\!+\!4A\mathcal{Q}f_k.
\end{split}
\end{equation}
Here $\mathcal{Q}=\sum\limits_{k'}N_{k'}$;
$\mathcal{P}=\sum\limits_{k'}f_{k'}$. Hereafter the evident
relations $\omega_k=\omega_{-k}$, $N_k=N_{-k}$ are used.

In the general case, Eqs. (\ref{eq:20}) are a set of nonlinear
integral equations. To simplify the situation, let us consider the
case of finite longitudinal size $L$, when only a single (two-fold
degenerate) level lies within the energy interval of about $2E_{0}$.
The level position $\xi$ with regard to the resonance center is
determined by the relation $\omega_{k}-\omega_0/2=\xi$. Let us
represent the function $f_k(t)$ in the form
$f_k(t)=|f_k(t)|\exp[i\varphi_k(t)]=N_k(t)\exp[i\varphi_k(t)]$
($N_k\gg 1$). Then, from Eqs. (\ref{eq:20}) we find
\begin{equation} \label{eq:21}
\begin{split}
&\frac{dN_{k}}{dt}=-2\gamma N_{k}+2E_0 N_{k}\cos\varphi_{k};\\
&\frac{d\varphi_{k}}{dt}=-2[\xi+\bar{A}N_{k}+E_0 \sin\varphi_{k}],\;\;\;\bar{A}=3A.
\end{split}
\end{equation}
The stationary solution of this system has the form
\begin{equation} \label{eq:22}
N_{k}^s=\frac{\sqrt{E_0^2-\gamma^2}\mp\xi}{|\bar{A}|};\;\;\;\;
\sin\varphi_{k}^s=\mp\frac{\sqrt{E_0^2-\gamma^2}}{E_0}.
\end{equation}
The signs $\mp$ correspond to $\bar{A}\gtrless0$, respectively. For
definiteness, we suppose $E_0>0$. The result (\ref{eq:22}) has an
interesting physical origin. In fact, the interaction $H_{eff}$
determines the effective renormalization of the phonon level,
leading to $\delta\omega_{k_0}=\bar{A}N_{k}$ for the case concerned.
Accordingly, $\xi\rightarrow \bar{\xi}=\xi+\bar{A}N_{k}$. From Eqs.
(\ref{eq:22}) it straightforwardly follows that
$\bar{\xi}_s=\pm\sqrt{E_0^2-\gamma^2}$. So, at $N_k=N_k^s$ the
renormalized level reaches the left or right  edge of the parametric
energy interval (within the accuracy of the shift due to $\gamma$).
As a result, the parametric increase of the phonon occupation
numbers stops and the phase acquires the constant value. The maximal
value of the phonon number equals
\begin{equation} \label{eq:23}
N_{kmax}^s\approx\frac{E_0}{|\bar{A}|}\gg 1;\;\;\;\;\gamma,\xi\ll
E_0.
\end{equation}
The phase is $\varphi_{k}^s\approx\mp\pi/2$ for $\bar{A}\gtrless
0$. The phase $\varphi_k$ corresponds to the phase correlation of
phonon pairs with the opposite momenta. The appearance of the
phase $\varphi_k$ and anomalous correlator $f_k$ is connected,
first of all, with the creation of phonon pairs in the course of
the evolution as a result of the parametric resonance. The
phonon-phonon interaction does not disturb the phase correlation,
and the evolution finishes in the stationary condensate of phonon
pairs with zero momentum and the common phase $\varphi_k^s$.

The character of the temporal evolution, tending asymptotically to
the values (\ref{eq:22}), (\ref{eq:23}), depends essentially on
the relations between the parameters in Eqs. (\ref{eq:21}). First,
let $\gamma=0$. In this case, Eqs. (\ref{eq:21}) have the
conserved integral of motion
\begin{equation}\label{eq:24}
H_0=2E_0 N_{k}\sin\varphi_{k}+ \bar{A}N_{k}^2+2\xi N_{k}.
\end{equation}
At the initial time moment when $\bar{\xi}_0=\xi+\bar{A}N_{k}(0)$
and $\bar{\alpha}_0=\sqrt{E_0^2-\bar{\xi}_0^{2}}$, Eqs.(\ref{eq:21})
imply that $\cos\varphi_{k}(0)=\bar{\alpha}_0/E_0$. Bearing in mind
that $\sin\varphi_{k}(0)=-\bar{\xi}_0/E_0$, from the second equation
in (\ref{eq:21}) we obtain $(d\varphi_{k}/dt)(0)=0$. As a
consequence, the initial (and conserved) value of $H_0$ is equal to
$H_0\approx -\bar{A}N_{k}^2(0)\ll E_0$. For an arbitrary time, Eq.
(\ref{eq:24}) can be rewritten as
$2E_0\sin\varphi_{k}(t)=-[2\xi+\bar{A}N_{k}(t)]-H_0/N_{k}(t)$.
Substituting this relation into equations (\ref{eq:21}), one can
find
\begin{equation} \label{eq:25}
\frac{dN_{k}}{dt}=\pm 2N_{k}\alpha;\;\;\;
\frac{d\varphi_{k}}{dt}=-\bar{A}N_{k}-\frac{H_0}{N_{k}},
\end{equation}
$\alpha=[E^2_0 -(\bar{A}N_{k}/2+\xi-H_0/2N_{k})^2]^{1/2}$. The signs
$\pm$ correspond to the regions with $|\varphi_{k}|\leq\pi/2$ and
$|\varphi_{k}|>\pi/2$, respectively. For the region with
$|\varphi_{k}|\leq\pi/2$, the solution can be found
straightforwardly
\begin{equation}\label{eq:27}
2t\approx \int\limits^{N_{k}(t)}_{N_{k}(0)}\frac{dx}{x[E^2_0
-(\bar{A}x/2+\xi)^2]^{1/2}}.
\end{equation}
Here we omitted the small term with $|H_0|\ll E_0$. The upper limit
of the integral is equal to the value $N_{k}^m\approx
2(E_0\mp\xi)/|\bar{A}|$ at which the denominator vanishes. The
divergence at $N_k\rightarrow 0$ is a typical manifestation of the
parametric resonance which requires the finiteness of the initial
field amplitude. This can be achieved by taking  zero-point
oscillations into account \cite{K1}. Owing to these oscillations, we
can put $N_{k}(0)\sim 1$. The time necessary for the system to
achieve the maximal value $N_{k}^m$ is equal to
\begin{equation}\label{eq:27a}
t_m\approx
\frac{1}{2\alpha_0}\ln\Big[\frac{4E_0}{\bar{A}}\cdot\Big(1-\frac{\xi^2}{E_0^2}\Big)\Big],
\end{equation}
We see that the argument of logarithm is much greater than unity
at $\xi<E_0$ and $E_0\gg \bar{A}$. This implies that $t_m\gg
1/2\alpha_0$ where $1/2\alpha_0$ is the characteristic time of the
parametric resonance. At $t=t_m$ we have $|\varphi_{k}|=\pi/2$. As
it follows from the second equation in (\ref{eq:25}), at $t>t_m$
the phase proves to be in the region with $|\varphi_{k}|>\pi/2$.
As a result, in the first equation the sign becomes negative and
the phonon number reduces. This is the start of an oscillating
behaviour. The joint solution of Eqs. (\ref{eq:25}) has the form
of high anharmonic oscillations of both the phonon number and the
phase around their stationary values. At $\xi=0$ the solution can
be written in terms of Jacobian elliptic functions. Averaging the
dependence $N_{k}(t)$ over the large temporal interval $t\gg t_m$,
we arrive at value $N_{k}^s$.

The character of evolution changes drastically at $\gamma\neq0$.
Namely, a self-averaging arises, leading asymptotically to the
stationary value $N_{k}^s$.
\begin{figure}[!h] \centering\vspace*{-0.5cm}
  \begin{minipage}[t]{.23\textwidth} \centering
    \includegraphics[angle=-90, width=\textwidth]{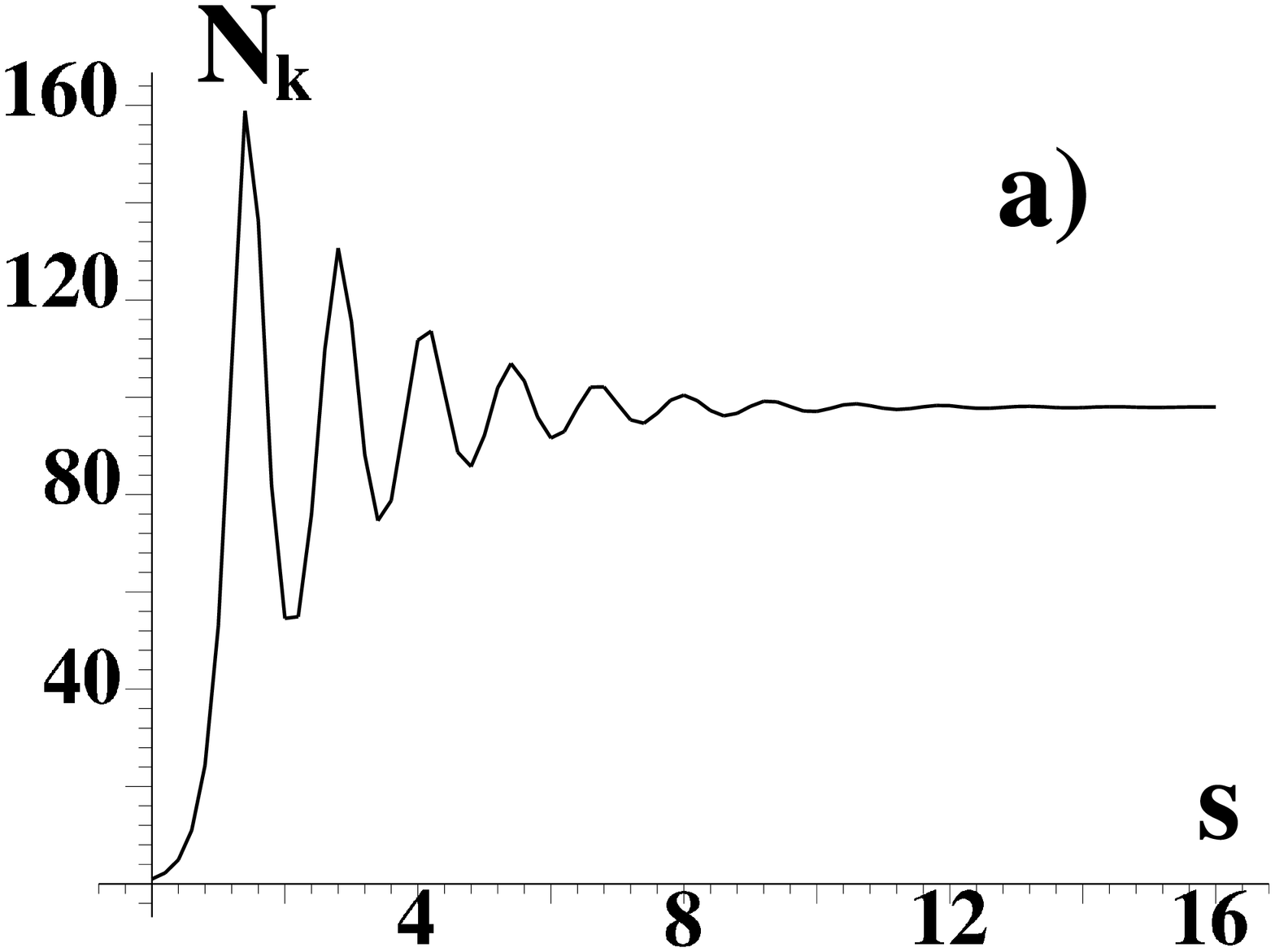}
\end{minipage}
  \begin{minipage}[b]{.23\textwidth} \centering
 \includegraphics[angle=-90, width=\textwidth]{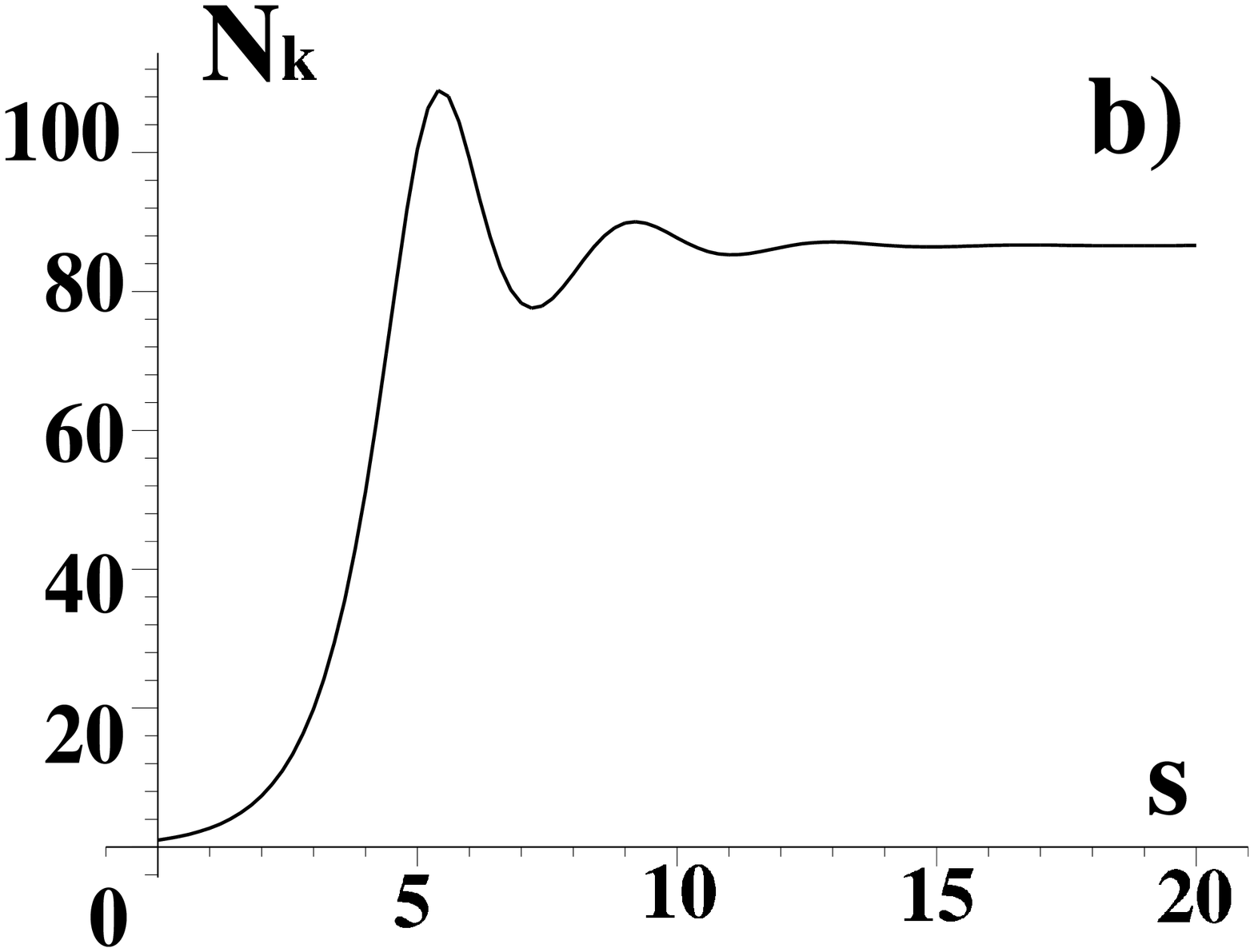}
\end{minipage}
\end{figure}
\begin{figure}[!h] \centering\vspace*{-0.7cm}
  \begin{minipage}[t]{.225\textwidth} \centering
    \includegraphics[angle=-90, width=\textwidth]{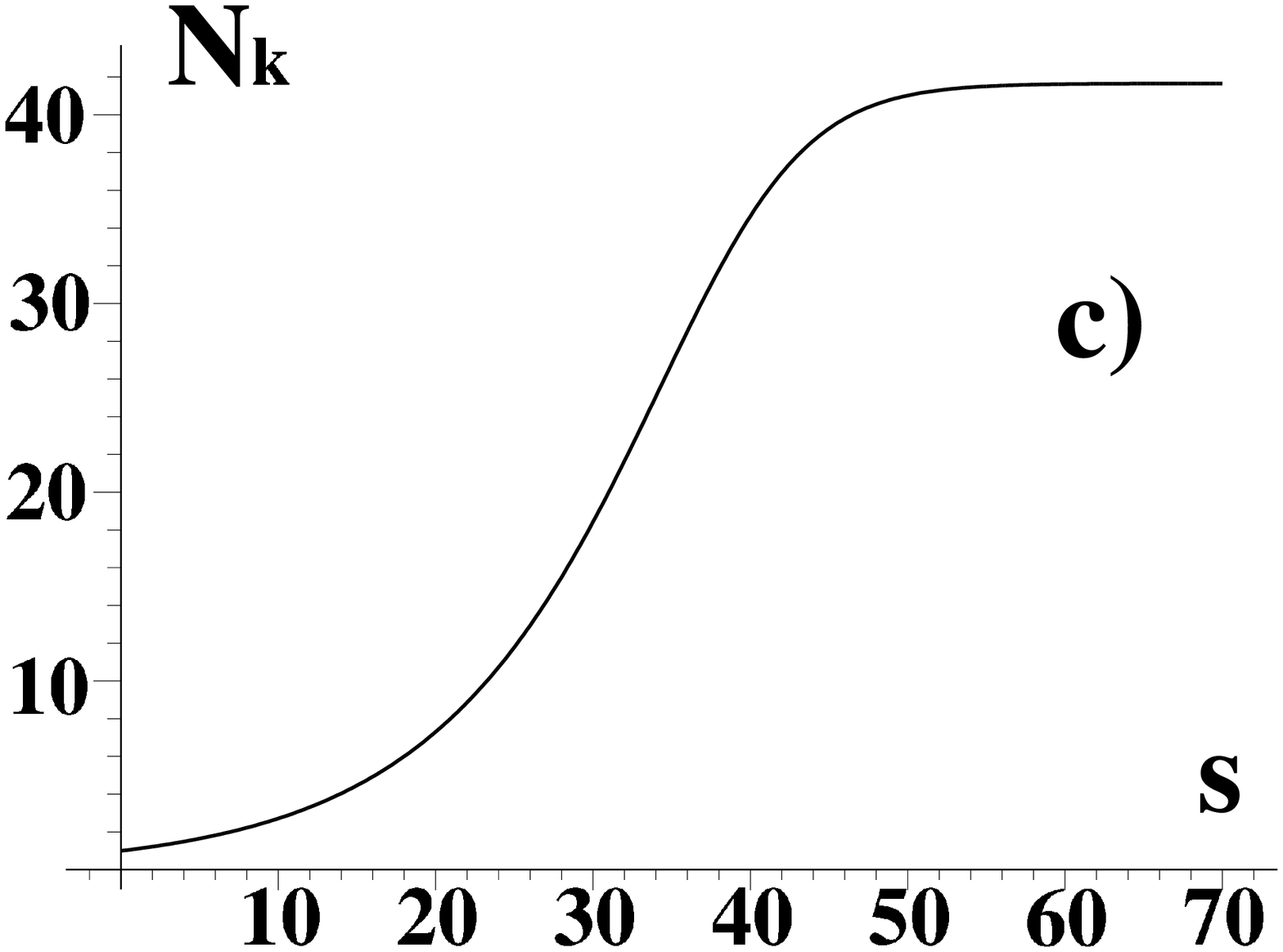}
\end{minipage}
  \begin{minipage}[b]{.225\textwidth} \centering
 \includegraphics[angle=-90, width=\textwidth]{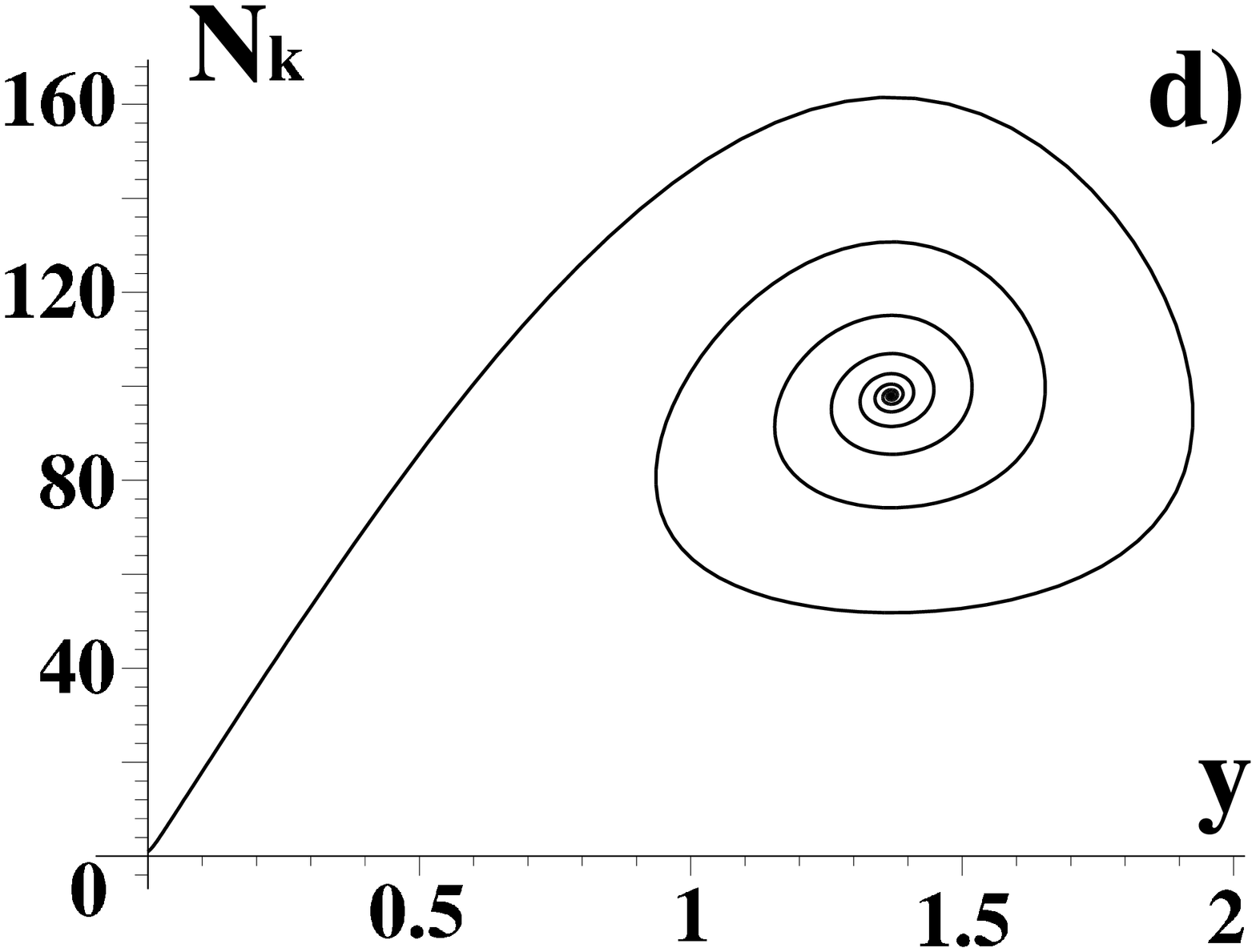}
\end{minipage}
\caption{! The phonon numbers $N_k(s)$ versus $s=2\gamma t$. The
cases (a), (b), and (c) for $(E_0/\gamma)=5,\;2,\;1.1$. The case
(d) plots $N_k(t)$ versus $y(t)=-\varphi_k(t)$ for
$(E_0/\gamma)=5$.}
\end{figure}
In this case, the integral of motion is absent and the solution of
Eqs.(\ref{eq:21}) should be found directly. The damping of the
oscillations has the decrement close to $\gamma$ when
$\alpha_0\gg\gamma$ and $t>t_m$. As $\gamma$ increases, the
arrival time to the stationary state decreases as $1/\gamma$. When
$\gamma$ is close to $\alpha_0$, the oscillations must disappear
completely. In fact, the time necessary for the phonon number to
reach the stationary state will be determined by Eq.
(\ref{eq:27a}) in which $\alpha_0$ is replaced by
$\alpha_1=\alpha_0-\gamma$. The direct numerical simulation of the
system (\ref{eq:21}) demonstrates the picture described. In Figs.1
(a)-(c), the phonon number $N_{k}$ as a function of $s=2\gamma t$
is shown at various values of the ratio $E_0/\gamma$ for the fixed
value of the ratio $(E_0/\bar{A})=10^2$ and $\xi=0$. As an
illustration, Figure 1 (d) displays the phase portrait in the
plane $[N_{k},\varphi_{k}]$ for $(E_0/\gamma)=5$. Thus the
self-consistent temporal evolution of interacting phonons near the
parametric resonance results in the formation of the stationary
state with the macroscopic phonon number (of the order of
$E_0/\bar{A}$) in a single quantum state. The stationary state has
one of two energies $\varepsilon_s=\hbar(\omega_0/2\pm E_0)$. In
the spatial sense this is a superposition of two states with the
wave vectors $k_s$ and $-k_s$.

The involvement of a coherent combination of phonons with the
momenta $k_s$ and $-k_s$ induces the stationary space modulation of
the condensate. It is easy to reveal determining the particle
density as $\Delta n=<\hat{\chi}^{'+}({\bf \rho},z)\hat{\chi}'({\bf
\rho},z)>$, where $\hat{\chi}'$ has the form (\ref{eq:11}).
Rewriting $\hat{\chi}'$ in terms of phonon operators and integrating
over $d^2 \rho$, we arrive at the expression for the modulation of
the 1D atomic density
\begin{equation} \label{eq:29}
\frac{\delta n(z)}{n^{(1)}}\!\!=\!\!\frac{2N^s_{k}}{N}(u_{k_s}^2+v_{k_s}^2)\cos 2k_s
z\!\!\approx\!\!\frac{2\mu}{\varepsilon_s}\cdot\frac{N_{k}^s}{N}\cdot\cos 2k_s z,
\end{equation}
where $n^{(1)}=N/L$. We suppose that $|\delta n|/n^{(1)}\ll 1$.

Let us make here some estimations. As an illustration, we consider
a quasi-1D gas of sodium atoms. The quasi-1D situation occurs
under condition $\mu<\omega_{\bot}$. Here $\mu\sim (a
n^{(1)})\hbar\omega_{\bot}$ (see \cite{O}), $a$ being the 3D
scattering length. For both $n^{(1)}=10^6 cm^{-1}$ and
$L\approx10^{-2}cm$, we have $an^{(1)}\approx 0.2$ and
$N\approx10^4$. Assuming $(\omega_{\bot}/\omega_0)=5$, we find
$$N_k^s\approx
10^4\cdot\frac{|\delta\omega_{\bot}|}{\omega_{\bot}};\;\;\;\frac{|\delta
n|}{n^{(1)}}\approx
\frac{3|\delta\omega_{\bot}|}{\omega_{\bot}}.$$ The above analysis
is performed for the case when a single level lies within the
energy interval $2E_0$. This assumption imposes certain
restrictions on the  ratio $E_0/\Delta\omega$ where
$\Delta\omega=\pi c/L$ is the spacing between the neighbor
longitudinal levels, $c$ being the sound velocity. Combining
definition of $E_0$ (\ref{eq:15}) and value  $\omega_{\bot}=10^4
1/sec$, we find $(E_0/\Delta\omega)\approx
10|\delta\omega_{\bot}|/\omega_{\bot})$. These estimations
demonstrate that at $|\delta\omega_{\bot}|/\omega_{\bot})\sim
10^{-2}$ the conditions $|\delta{n}|/n^{(1)}\ll 1$,
$E_0/\Delta\omega\ll 1$ are satisfied and  value $N_k^s$ proves to
be about $10^2$. In reality, we may increase the ratio
$(|\delta\omega_{\bot}|/\omega_{\bot})$. As a consequence, the
value $N_k^s$ should be within  interval $10^2-10^3$. The
parameter $\gamma/E_0$ essential for the temporal evolution may be
changed varying the ratio $|\delta\omega_{\bot}|/\omega_{\bot}$.
Thus, we can conclude that the phonon condensation can be realized
for a quite realistic set of parameters. It should be noted that,
for the finite but sufficiently low temperatures when the initial
number of phonons with $\omega_k\sim\omega_0$ is limited, the
results obtained remain valid.


\begin{thebibliography}{99}
\bibitem{K1} Yu.Kagan and L.A.Maksimov, Phys.Rev.A, {\bf 64}, 053610 (2001); Laser
Physics, v.12, p.153 (2002).
\bibitem{Ch} F.Chevy, V.Bretin, P.Rosenbusch, K.W.Madison, and J.Dalibard,
Phys.Rev.Lett., {\bf 88}, 250402 (2002).
\bibitem{K2} Yu.Kagan and L.A.Maksimov,Phys.Lett.A,{\bf 317},477(2003).
\bibitem{K3} Yu.Kagan, E.L.Surkov, and G.V.Schlyapnikov, Phys. Rev. A, {\bf 54}, R1753 (1998).
\bibitem{Z} E.Zaremba, Phys.Rev.A, {\bf 57}, 518 (1998); S.Stringari,  Phys. Rev. A, {\bf 58}, 2385 (1998).
\bibitem{P} Fr.Dalfovo, St.Giorgini, L.P.Pitaevskii, and S.Stringari, Rev.Mod.Phys., {\bf
71}, 463 (1999).
\bibitem{L} E.M.Lifshitz and L.P.Pitaevskii, Statistical Physics (Pergamon, Oxford, 1980), Part 2.
\bibitem{O} M.Olshanii, Phys.Rev.Lett., {\bf 81}, 938 (1998).
\end{thebibliography}
\end{document}